\documentclass[12pt]{iopart}
\usepackage{epsfig}
\begin{document}

\title{Sympathetic ground state cooling and coherent manipulation with two-ion-crystals}

\author{H. Rohde, S. T. Gulde, C. F. Roos, P. A. Barton, D. Leibfried,
J. Eschner, F. Schmidt-Kaler \footnote[3]{e-mail:
Ferdinand.Schmidt-Kaler@uibk.ac.at}, and R. Blatt}

\address{Institut f\"ur Experimentalphysik, Universit\"at Innsbruck, Technikerstrasse 25, A-6020 Innsbruck, Austria}

\begin{abstract}
We have cooled a two-ion-crystal to the ground state of its
collective modes of motion. Laser cooling, more specific resolved
sideband cooling is performed sympathetically by illuminating only
one of the two $^{40}$Ca$^+$ ions in the crystal. The heating
rates of the motional modes of the crystal in our linear trap have
been measured, and we found them considerably smaller than those
previously reported by Q. Turchette {\em et. al.} Phys. Rev. A 61,
063418 (2000) in the case of trapped $^9$Be$^+$ ions. After the
ground state is prepared, coherent quantum state manipulation of
the atomic population can be performed. Within the coherence time,
up to 12 Rabi oscillations are observed, showing that many
coherent manipulations can be achieved. Coherent excitation of
each ion individually and ground state cooling are important tools
for the realization of quantum information processing in ion
traps.
\end{abstract}

%\pacs{32.80.Pj, 03.67.Lx, 42.50.Lc}
\date{today}

\section{Introduction}

The coherent control of quantum systems has been given much
attention and experimental effort in recent years. In particular,
there has been a concerted effort in developing the experimental
tools needed to create scalable quantum information processing
(QI), and in smaller `proof of principle' demonstrations of QI.
The use of the quantum systems at the heart of these experiments
relies on the entanglement between subsystems. This might be
easily destroyed by the coupling to the environment. As a
suggestion for a well controllable, almost decoherence-free
system with which to realise a quantum computer, a linear ion
trap was proposed by Cirac and Zoller~\cite{C-Z}. Indeed, this
idea seems so promising at present that many groups are actively
pursuing research towards quantum state preparation and
manipulation in ion
traps.~\cite{ent4ions,spectrosc,stateeng,lalamos,oxford,waltchain}

The Cirac \& Zoller scheme uses the internal states of a chain of
trapped ions as qubits. Each of these ions represents a qubit and
needs to be manipulated separately and distinctly with laser
light~\cite{address}. Ions in the chain are then coupled by the
laser excitation of vibrational modes of the crystal within the
trapping potential. As this quantum of motion is common to all of
the ions, it acts as a so called 'quantum bus' (gate mode),
transporting information in the chain of ions. In order to
implement the Cirac \& Zoller proposal for a quantum gate, the
gate mode must be prepared in the ground state prior to the gate
operation. Cooling of ion crystals has been previously
demonstrated \cite{cool2ions}. This technique is not only
important for ion trap QI's but also in frequency measurements on
chains of ions~\cite{freqmeas} and in QED experiments with trapped
ions.

So far, cooling to the motional ground state has been performed
only on $^9$Be$^+$ two-ion-crystals by illuminating the entire ion
chain~\cite{ent4ions,cool2ions}. In our experiment, we have
pre-cooled two-ion-crystals into the mK range by laser cooling
(Doppler cooling) and identified all motional eigenmodes. Then, we
have cooled the two-ion-crystal to the vibrational ground state by
illuminating only one of the two ions (sideband cooling). Thus we
present here the first demonstration of sympathetic ground state
cooling of an ion string. If we cool only {\em one of the
eigenmodes}, we achieve a ground state population of any of the
modes which is greater than 95$\%$. The other modes are left in
their thermal states. However, the fidelity of coherent
manipulation of the ion's quantum state would be spoiled due to
the thermal distribution of the other modes.

Sequentially we have cooled {\em all eigenmodes} and we achieve a
greater than 95$\%$ ground state population of the axial mode
(gate mode) and a phonon number between 1 and 2 for the other
modes ('spectator' modes). After that, coherent manipulation of
the ions' state is possible. We choose two levels, the $S_{1/2}$
ground state and the long lived $D_{5/2}$ metastable state, for
the qubit basis for QI processing. In our experiment, we
demonstrate how each of the ions'  state can manipulated (''single
qubit rotation''). The coherence of this optical state
manipulation is shown by 12~Rabi oscillations with better than
60\% contrast.

We measure heating rates of the vibrational eigenmodes and find
they are all well below 20 phonons~s$^{-1}$. This slow heating
rate, in combination with the sympathetic cooling method, will be
of importance for QI:  Recent proposals suggesting continuous
sympathetic ground state cooling together with coherent
manipulation of qubits at different locations in the ion string
\cite{othersymcool,waltscool} appear no longer unrealistic in view
of these findings.

\begin{figure}
\begin{center}
\epsfig{file=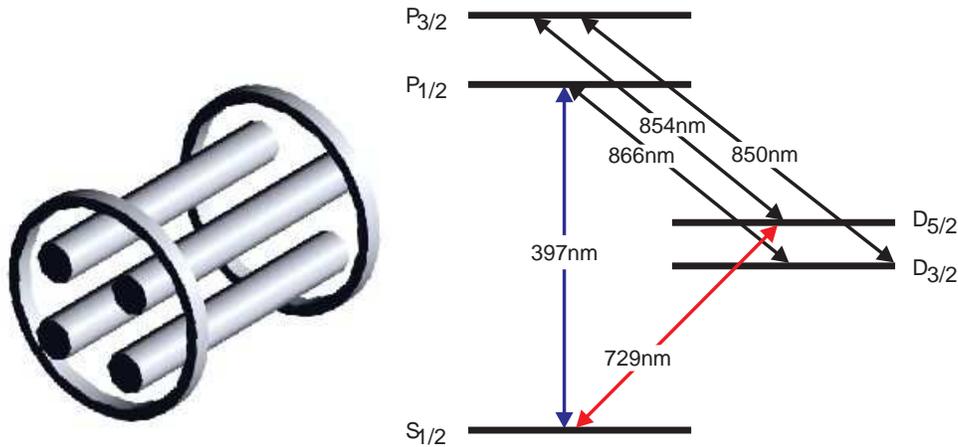, height=6cm} \caption{Left: A figure of the
trap showing the ring endcaps and the four rf electrodes. Right:
$^{40}$Ca$^+$ level scheme. Only relevant levels
 and their transition
wavelengths are shown. } \label{f_levels} \label{f_trap}
\end{center}
\end{figure}

\section{The experiment}
% trap:
In our experiments we confine $^{40}$Ca$^+$ ions in a linear Paul
trap~\cite{spectrosc}. Four  stainless steel rods  of 0.6~mm
diameter are arranged parallel at the corners of a square shape of
side of length 2.1~mm. The direction of the rods defines the axial
symmetry of the trap (fig. \ref{f_trap}, left). An alternating
radio frequency (rf) field is applied to two diagonally located
rods while  both the other rods are held at zero volts. In such a
way, an electric quadrupole field along the axial direction is
generated which provides radial confinement for a charged particle
if the rf-frequency and amplitude is chosen properly \cite{Paul}.
Two stainless steel endcap rings of inner diameter 6.7~mm spaced
10~mm apart over the rods are charged with a positive voltage
which provides axial confinement (fig.~\ref{f_trap}, left). The
electrode arangement is held together using isolating macor
spacers (not shown in the figure) which assure a 20~$\mu$m
tolerance in the position of the four rods and the endcap
electrodes. The trap is housed in a vacuum chamber with a pressure
less than 10$^{-10}$~mbar.

The rf voltage applied to the rods is produced by resonantly
enhancing a source ($\omega_{rf}$~=~16.0~MHz and typical power
4~W) in a helical resonator (loaded Q~$\approx$~250). These
parameters lead to an rf voltage of about 300~V$_{rms}$ and
typical radial secular frequencies of $\omega_{rad}$~=~1.8~MHz.
The radial x- and y-directions (orthogonal to the axial direction)
are degenerate in frequency. For axial confinement, we apply  a DC
voltage to the ring shaped endcap electrodes (fig.~\ref{f_trap},
left). For the typical trapping operation we use 2~kV which gives
an axial secular frequency of $\omega_{ax}$~=~700~kHz. With this
ratio of radial and axial frequency, crystals with up to 5~ions
arrange themselves in a linear string. Ions are generated by
electron bombardment of a weak neutral Ca atomic beam, and loaded
into the trap. This loading process also generates electric stray
voltages, shifting the ions out of the trap centre, which is the
rf field node line, and resulting in a driven 'micro'-motion of
the chain. We compensate for the stray charges using the following
procedure: With laser beams at 397~nm from different directions
the residual micromotion is detected in the respective directions
via the modulation of the fluorescence rate, induced by the
Doppler shift of the moving ions \cite{BERKELAND98} and the
micromotion is nulled by the application of well chosen voltages
on additional compensation electrodes. Due to the slight
asymmetries of the trap construction, a compensation voltage
dependent on the endcap DC voltage is required. In addition, due
to the stray charges (possibly on the macor surfaces of the trap
arrangement), we need compensation voltages which vary from load
to load. Furthermore, these stray charges leak away within a
timescale of hours which requires frequent recompensation. Once
compensated, linear ion string configurations lie on the rf node
line, thus having no micromotion, which is of great importance for
cooling (see sect. 3) and coherent manipulation of the qubit
transition, as far as QI processing is concerned.

%two-ion cystal modes:
Since a single trapped particle has three normal modes of motion,
the two-ion-crystal vibration is described by 6 vibrational modes
\cite{James,Steane} : Three centre of mass modes (of which one is
axial and two are radial) at $\omega_{ax}$ and
$\omega_{rad}^{(x,y)}$ respectively, two 'rocking' modes  at
$\omega_R^{(x,y)}$~=~$(\omega_{rad}^{(x,y) \hspace{0.1cm} 2} -
\omega_{ax} \hspace{0.1cm} ^2)^{1/2}$ (where the atoms swing in
opposite directions radially) and the 'breathing' mode (where the
two ions oscillate exactly out of phase along the axial direction)
at $\omega_b$~=~$\sqrt{3}~\omega_{ax}$. In the specific case of
this trap, the radial x- and y-oscillation frequencies are
degenerate, giving rise to four distinct normal mode frequencies
in total.

%ca levels, lasers, detection method:
We trap $^{40}$Ca$^+$ ions (relevant energy levels shown in
figure~\ref{f_trap}, right). All the necessary wavelengths of
light can be generated using solid state lasers and frequency
doubling techniques~\cite {spectrosc}. Light at a wavelength of
397~nm is produced by frequency doubling a stabilized Ti:Sapphire
laser (linewidth $<$~300~kHz). Using the $S_{1/2} \rightarrow
P_{1/2}$ transition, we perform conventional Doppler cooling and
detect the ions resonance fluorescence on an intensified CCD
camera and a photomultiplier. The $P_{1/2}$ state may decay with a
branching ratio of 1/15 into the metastable $D_{3/2}$ state, so a
diode laser at 866~nm is used to repump the population back via
the $P_{1/2}$ state. A second diode laser at 854nm can be used to
remove the population in the $D_{5/2}$ state via a transition to
the $P_{3/2}$ state. The quadrupole transition near 729~nm
connects the $S_{1/2}$ electronic ground state to the metastable
$D_{5/2}$ state (lifetime $\sim$ 1~s). The S$_{1/2}$ to D$_{5/2}$
quadrupole transition at 729~nm is excited with a highly stable
Ti:Sapphire laser (linewidth $<$~100~Hz). This transition is used
for the implementation of the two logic states of the qubit: We
can detect whether a transition to D$_{5/2}$ occurred by applying
the beams at 397~nm and 866~nm and monitoring the fluorescence of
the ion. In 12~ms we typically collect 30~fluorescence photons per
ion on a stray light background of 10~photons if the ion is in the
electronic ground state \cite{coat}. The internal state of the ion
can thus be discriminated with an efficiency above 98\%
\cite{christian}.

\section{Doppler cooling \label{doppl}}
After Doppler cooling, the ion cloud crystallizes into a linear
string, thus performing the initial cooling step towards the
motional ground state \cite{Wineland79}.  Optimum Doppler cooling
is achieved at a laser frequency of $\delta \omega = - \Gamma /2
$, red detuned from the resonance, and leads to a temperature in
the mK regime, with the natural linewidth of the $ S_{1/2}
\rightarrow P_{1/2} $ transition being $ \Gamma $= 20~MHz. In the
case of one cooling laser beam, with optimum detuning and incident
parallel to the oscillator axis, the Doppler cooling limit is
given as $ k_B T = \frac{1}{2} \hbar \Gamma $.

In our specific case, at least a number of additions to the above
simple picture affect the cooling limit: (i)~$^{40}Ca^+$ is not a
true two-level-system since both the $S_{1/2} \rightarrow P_{1/2}$
at 397~nm and the $D_{3/2} \rightarrow P_{1/2}$ transition at
866~nm have to be excited for Dopper cooling. As shown, for
example, in ref. \cite{REISS96} this leads to coherent
superposition states of the $S_{1/2}$ and the $D_{3/2}$ state
('dark resonances'), which can increase or decrease the Doppler
cooling limit, depending on the relative detuning of both laser
frequencies. Since dark resonances can exhibit spectral features
much narrower than the natural linewidth, the detuning of the
lasers strongly influences the cooling. (ii)~We apply a magnetic
field of 3.6~Gauss (to generate a quantization axis), which leads
to a frequency splitting of all Zeeman substates. With the two
laser beams at 397~nm we simultaneously drive  $\pi$-transitions
and $\sigma^- \hspace{0.05cm} / \hspace{0.05cm} \sigma^+$
-transitions. Thus, four resonances between the $S_{1/2}, m_J= \pm
1/2 $ manifold and $ P_{1/2}, m_J= \pm 1/2 $ manifold are excited,
all of them slightly shifted against each other in frequency due
to different Land$\acute{e}$ factors, which leads to an effective
broadening of the cooling transition. For example, the $ \sigma^-
$ and the $\sigma^+$ -transitions are split by $\approx$~13~MHz.
The laser detuning can no longer be optimally chosen for Doppler
cooling with respect to {\em all transitions} between the Zeeman
substates $ m_J= \pm $ 1/2. (iii)~Residual micromotion of the
trapped ions leads to a larger cooling limit. With a rf drive
frequency of 16~MHz, the micromotion modulates the bare spectral
line, generating resonances ('sidebands') which are not resolved
under the natural linewidth of the $S_{1/2} \rightarrow P_{1/2} $
transition. These unresolved micromotion components broaden the
cooling transition. (iv)~We try carefully to avoid any saturation
of the cooling transition which would lead to further broadening.

Due to the combination of reasons (i) to (iv), we can explain the
somewhat larger width of the observed excitation spectrum on the
$S_{1/2} \rightarrow P_{1/2} $ transition of $\Gamma_{exp} \approx
$~30~MHz. (v)~It seems that intrinsic heating processes do not
play any role if we assume the measured heating rates for the
ground state (see section \ref{cooling}) to be approximately valid
also at higher phonon numbers, $\bar{n}$. (vi)~The limit of
Doppler cooling can always be expressed as the balance between
cooling processes and heating processes. If the laser beam
direction is not parallel with the oscillator axis, but has an
angle with respect to the oscillator axis, the probability for the
cooling process decreases, while the heating process (by
spontaneous emission) is independent of the beam direction. As a
result, the cooling limit is increased: For our experimental
arrangement with two laser beams, and taking into account the
angles between these cooling beams and the principal trap axes, we
estimate that in the axial motional mode a cooling is reached
which corresponds to a mean phonon number of $\bar{n}_{axial}$ =
45(10) and $ \bar{n}_{radial}$ = 40(10). For the calculation of
these phonon numbers, we have taken the effective linewidth of
30~MHz into account. The uncertainty is caused by imprecise
knowledge of the exact fraction of light power going to each of
the cooling beams and their exact focus sizes. Spectroscopy on the
$S_{1/2} \rightarrow D_{5/2}$ transition (as will be discussed in
the next section) yields 50(10) in the radial, and 45(10) phonons
in the axial direction.

\begin{figure}
\begin{center}
\epsfig{file=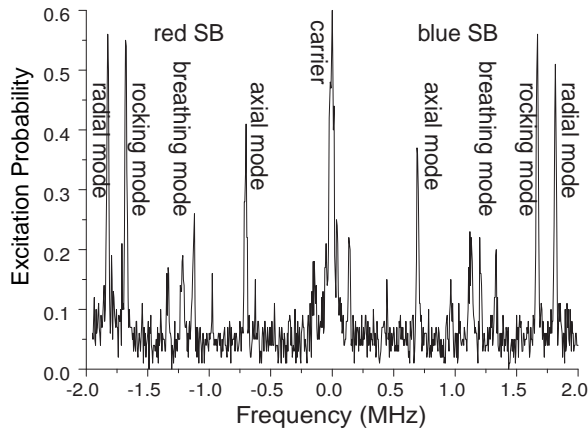, height=6cm} \caption{Spectrum of a two ion
chain, excitation probability is plotted against 729~nm laser
detuning. The detuning of the laser is taken to be zero at the
carrier frequency. All first order sidebands are named as in the
text.} \label{f_spec}
\end{center}
\end{figure}

%laser spectroscopy at 729 principle:
\section{Spectroscopy of the $S_{1/2}$ - $D_{5/2}$ transition \label{spec}}
For coherent spectroscopic investigation and state engineering on
the S$_{1/2} \leftrightarrow$ $D_{5/2}$ transition at 729~nm we
use a pulsed technique which consists of four consecutive steps.
(i)~{\em Doppler cooling}: Laser light at 397~nm, 866~nm, and
854~nm is used to pump the ion to the $S_{1/2}$ ground state and
pre-cool the vibrational state to a value near the Doppler limit.
(ii)~{\em Optical pumping}: The S$_{1/2}(m_J=-1/2)$ sub-state is
prepared by optical pumping with $\sigma^-$ radiation at 397~nm. A
magnetic field of 3.6~Gauss at right angles  to the direction of
the $k$-vector of the light at 729~nm provides a quantization axis
and splits the 10~Zeeman components of the $S_{1/2}
\leftrightarrow$ $D_{5/2}$ transition in frequency. (iii)~{\em
Excitation step:} We excite the $S_{1/2}(m_J=-1/2)
\leftrightarrow$ $D_{5/2}(m=-5/2)$ transition at 729~nm with laser
pulses of well controlled frequency, power, and duration.
(iv)~{\em State analysis:} We collect the ion's fluorescence under
excitation with laser light at 397~nm and 866~nm and detect
whether a transition to the shelving level $D_{5/2}$ has been
previously induced.

Sequence (i)-(iv) is repeated 100 times to measure the $D_{5/2}$
state occupation probability $P_D$ after step (iii). We study the
dependence of $P_D$ on the experimental parameters such as the
detuning $\delta\omega$ of the light at 729~nm with respect to the
ionic transition or the length of one of the excitation pulses in
step (iii). The duration of a single sequence is typically 20~ms
(in some cases 40~ms), so that we can synchronize the sequence
with the ac power line at 50~Hz to reduce the influence of
ac~magnetic field fluctuations.

The direction of the laser at 729~nm, which is used for
spectroscopy, resolved sideband cooling and for individual quantum
manipulation, is at 67.5~degrees with respect to the axial
direction. The ion-light interaction \cite{ionlight} is described
by the Lamb Dicke parameters $\eta_{ax}=4.3\% $,
$\eta_{rad}=6.6\% $ for the excitation of the axial and radial
sidebands respectively ($\eta$ being defined as $k \cos \phi
\sqrt{\hbar / 2m \omega}$, where $\phi$ is the angle between the
oscillator axis and the incident laser beam, $m$ is the mass of
the ion crystal and $\omega$ is the relevant trap oscillator
frequency).

The observed excitation spectrum on the $S_{1/2}(m_J=-1/2)
\leftrightarrow$ $D_{5/2}(m_J=-5/2)$ transition (figure
\ref{f_spec}) consists of the following spectral resonances: First
there is an atomic line at the bare frequency of the transition
('carrier'). In addition to this line, there are spectral
resonances which are due to the harmonic oscillator frequencies of
the ions in the trap. Thus, the 4 distinct frequencies of motion
lead to 4 distinct resonances on the red wavelength side of the
central line and symmetrically  on the blue wavelength side. In
addition to these first order modulations, there are second order
modulations of the first order resonances themselves and so forth.
This results in a complicated comb structure of sidebands
\cite{spectrosc}. The  width of the comb is dependent on the
temperature of the ion chain. With the initial Doppler cooling
step we are able to reduce the thermal phonon number in all modes
of the two-ion-crystal sufficiently that only a few resonances are
strong, a typical spectrum is shown in figure \ref{f_spec} with
all first order sidebands named. We denote these resonances as
follows: The 'carrier' transition means the phonon number of all
motional modes remains the same when the ion undergoes the
transition from the $S_{1/2}$ ground state to the $D_{5/2}$ state.
A red (respectively blue) sideband means that the excitation of
the particular mode removes (adds) one phonon when the electronic
transition is excited.

%measurement of the doppler cooling limits:
Using the calculated Lamb-Dicke parameters in the radial and axial
directions, we compare the excitation dynamics on the $S_{1/2} -
D_{5/2}$ transition for the vibrational sidebands, quantified by
the population $P_{D5/2}(t)$ after an excitation time $t$, with
that on the carrier transition. For a thermal distribution with
mean phonon number $\bar{n}$, which is characterized by the phonon
distribution $p_n$, we expect a blue sideband Rabi oscillation of
$P_{D5/2}^{SB}(t) \propto \sum p_n \sin^2(\eta \sqrt{n+1} \cdot
t)$, while for the carrier transition the Rabi oscillation is
given by $P_{D5/2}^{Carrier}(t) \propto \sum p_n \sin^2((1- \eta^2
n) \cdot t)$ \cite{ionlight}. Thus, from the observation of Rabi
oscillations of $P_{D5/2}(t)$ we are able to deduce the mean
phonon number, $\bar{n}$, after Doppler cooling (section
\ref{doppl}). We typically find for a single Doppler-cooled
trapped ion, mean phonon numbers of radial $\bar{n}=50(10)$, and
axial $\bar{n}=45(10)$, which indicates that the ions are confined
within the Lamb-Dicke regime, with $\eta \sqrt{\bar{n}} < 1$.

\section{Sideband cooling of the two-ion-crystal}
\label{sideband} After Doppler cooling the two-ion-crystal into
the Lamb-Dicke regime, {\em resolved sideband cooling} is applied
in order to cool to the motional ground state. The sideband
cooling uses the quadrupole  S$_{1/2}(m_J=-1/2) \leftrightarrow$
D$_{5/2}(m_J=-5/2)$ transition which has a natural linewidth of
0.16~Hz. Only one of the ions is illuminated, and so only the
electronic population of one ion is driven although the motional
modes are common to both ions. First, the required frequencies
have already been determined by spectroscopic investigation as
described in section \ref{spec}. Then the laser at 729~nm is tuned
to the red sideband of the mode of motion we wish to cool. With
each photon absorbed, the ion-crystal loses one phonon of motional
energy of the specific mode. As the lifetime of the $D_{5/2}$
state is long, the population must be recycled quickly to the
lower state, thus the laser at 854~nm is switched on during this
sideband cooling pulse, quenching the atomic population in the
$D_{5/2}$ state via the $P_{3/2}$ state. Subsequent spontaneous
decay closes the cooling cycle, but does not change the phonon
number as the Lamb-Dicke parameter $\eta_{393}$ is small. The
approximate laser power between 5 to 20~mW at 729~nm is focused to
a waist size of 3.7~$\mu$m, and the power of the 854~nm (laser
focus size: 100 to 200~$\mu$m) is adjusted for optimum cooling,
typically a few tenths of a mW. The ion is illuminated with both
laser fields for a time of 6~ms. Under these operating conditions,
we observe efficient red sideband cooling at 729~nm within a
bandwidth of 5~kHz. During the cooling and quenching pulse, short
pulses of $\sigma^-$ polarised light are applied. With that, any
accidental loss of population into the $S_{1/2}(m_J=+1/2)$ state
is repumped and the ion is returned to the the cooling cycle. The
duration of these pulses is kept at a minimum to prevent unwanted
heating.

At the end of the cooling phase there is a probe pulse of light at
729~nm followed by detection on the Doppler cooling transition to
obtain the motional quantum state. Thus the ground state cooling
step is placed between steps (ii) and (iii) of the spectroscopic
sequence. The frequency of the 729~nm probe pulse is varied and
the spectrum after sideband cooling recorded. Any population in
the ground state of a motional mode will not be excited by the
probe pulse when tuned to the red sideband and so the strength of
this line in the spectrum will be greatly reduced. Measuring the
absorption on the red ($rsb$) and blue ($bsb$) sidebands of the
cooled mode allows us to determine the ground state population,
$p_0 = 1-(rsb/bsb)$.

\begin{figure}
\begin{center}
\epsfig{file=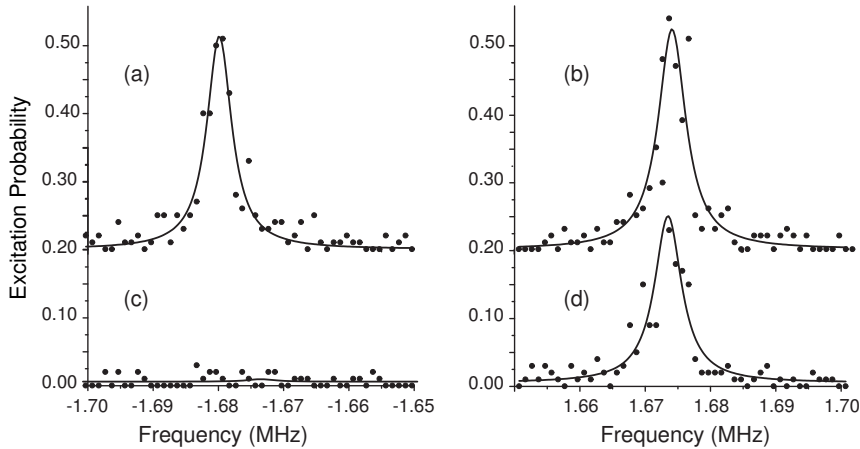, height=6cm} \caption{Sideband cooling on
the  $S_{1/2}$ - $D_{5/2}$ transition for two ions in the linear
trap. The radial sideband excitation is shown after Doppler
cooling (upper traces a, and b) and after ground state cooling
(lower traces c, and d). The y-axis of the upper traces has been
offset by 0.2 for clarity. From the asymmetry of red (a, c) and
blue sideband excitation (b, d) we deduce a 98.5(1.5)\% ground
state probability.} \label{f_radkalt}
\end{center}
\end{figure}

% one mode:
For cooling only a {\em single mode of the motion}, one cooling
pulse of 6.4~ms is used. This cooling time is chosen such that the
mode is cooled to the ground state. We have cooled all of the
motional modes separately using the scheme as described. The
obtained cooling results show that we are able to reliably cool
all modes separately to a ground state population of over 95\%.
Figure \ref{f_radkalt} shows the  red and blue radial sidebands
before and after cooling giving a 98.5(1.5)\% ground state
probabilty. We are able to prepare the 1-D zero point of motion of
any oscillator that could be used as the 'gate mode' in a
realisation of a small QI~processor. However, any uncooled modes
with thermal phonon number distributions ('spectator' modes),
affect the coherent dynamics on the cooled mode. An approximate
expression for the maximum number $N^*$ of Rabi cycles, is given
by $N^* \le 1/(2\eta^2 \bar{n})$, where $\bar{n}$ and $\eta$
accounts for the mean phonon number and Lamb-Dicke factor of a
spectator mode \cite{christian}. Thus if the average number of
phonons is high in a mode, significantly different Rabi
frequencies occur and  Rabi oscillations are washed out.
Therefore, in order to maximise the possible number of gate
operations these spectator modes must also be cooled.

\begin{figure}
\begin{center}
\epsfig{file=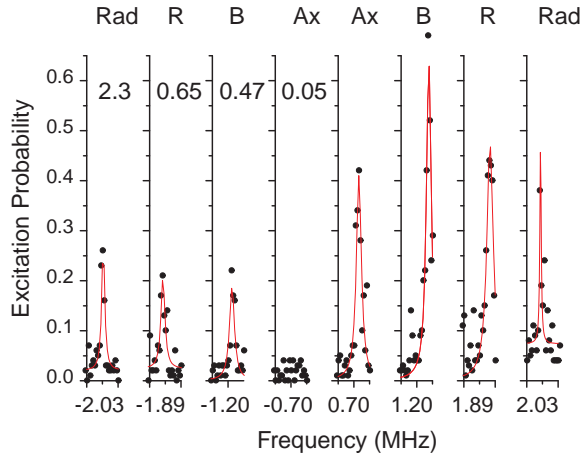, height=6cm} \caption{Four sideband
cooling, excitation probability is plotted against 729~nm laser
detuning. The four red and blue motional sidebands are shown after
cooling. The points are experimental data with fits to the data
shown as solid lines. The residual average phonon numbers in each
motional mode are shown. The final ground state population for the
axial mode is greater than 95\%.} \label{f_cool}
\end{center}
\end{figure}

%all modes
In order to cool {\em all of the motional modes} such that they
are all prepared close to the ground state, we use a scheme almost
identical to that above, although during the cooling pulse we
clearly need to apply light at all the red sideband frequencies.
So, rather than using a single cooling pulse we apply four
separate pulses, each 6~ms long, and each tuned to the red
sideband of a different oscillator, i.e. radial, rocking,
breathing and axial. During each cooling pulse the quenching and
repumping lasers are also applied, and once again only one ion is
illuminated. The power of the cooling laser at 729~nm of each
pulse is chosen to optimise cooling results.  Cooling is applied
in the following order: radial, rocking, breathing and axial.
Since we intend to use the axial motional mode as the gate mode,
this is the last to be cooled (with the lowest residual phonon
number). After cooling, the population is once again probed, and
the spectrum of all four motional modes is recorded. In
figure~\ref{f_cool} the red and blue sidebands of all four
motional modes are shown after cooling. In stark contrast to the
sidebands before cooling (in figure~\ref{f_spec} the red and blue
sidebands are of approximately the same height) those on the red
side have all but disappeared. For the scan shown in
figure~\ref{f_cool} this reveals the following average phonon
numbers, $\bar{n}=p_0^{-1}-1$, for each mode: radial
$\bar{n}_{rad}$~=~2.3, rocking $\bar{n}_{R}$~=~0.65, breathing
$\bar{n}_b$~=~0.47 and axial $\bar{n}_{ax}$~=~0.05.

As follows from the above discussion, we see that (i) the axial
gate mode phonon number is indeed well prepared for the
implementaltion of a cold gate, e.g. of Cirac \& Zoller type, and
(ii) the low phonon number of the spectator modes allows for high
contrast Rabi cycles since $\eta^2 \bar{n} \ll 1$.

\begin{figure}[b]
\begin{center}
\epsfig{file=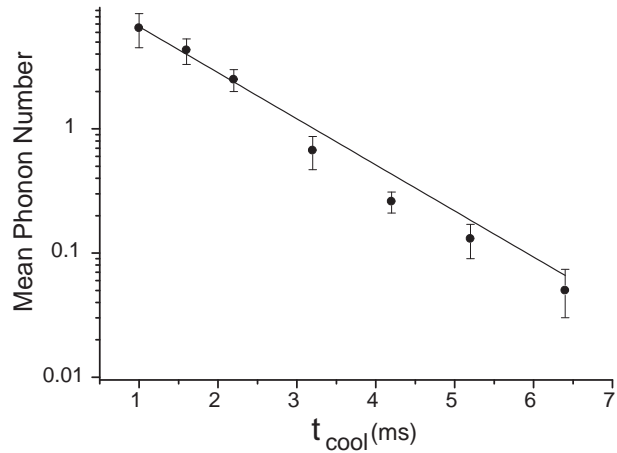, height=6cm} \caption{Cooling rate of
the rocking mode, phonon number is plotted against the duration of
the sideband cooling pulse. Points are experimental data with
error bars and the line is a fit to the data. The $1/e$ cooling
time is 1.2~ms and over 95\% ground state population is acheived
at the end of the cooling pulse}  \label{f_coolrate}
\end{center}
\end{figure}

\begin{table}[t]
\label{t_cool}
\begin{tabular}{l|ccccc}
Trap & Mode & Frequency $(MHz)$ & 1/e time (ms) & to 1 phonon (ms)
\\
  \hline Linear  & breathing & 1.2 & 3.2 & 4.0  \\
         Linear  & rocking & 1.7 & 1.2 & 3.0  \\
         Spherical, single ion & axial & 4 & 0.2 & $\le$ 0.2 & \\
         Spherical, single ion  & radial & 1.9 &  &  & \\
\end{tabular}
\caption{Cooling times for various motional modes, for the
spherical trap see \cite{christian,stateeng}.}
\end{table}

\section{Cooling times and heating rates}
%cooling and heating rates
%cooling rate first
Aside from the purity of preparation of the motional ground state,
the dynamics of cooling and heating is important. We thus
investigate the cooling  and heating rates of the various modes of
motion. For the cooling rate, we vary the cooling time,
$t_{cool}$, and measure the residual phonon number of the
respective mode as described earlier for the sideband cooling. The
remaining phonon number in the mode is then plotted against the
cooling time and characteristic times can be deduced. The result
is given as the time to reach $1/e$ of the initial phonon number
(see figure \ref{f_coolrate} and table~1). A second measure,
mainly important for the spectator modes, is the time which is
necessary for reaching a mean phonon number near one,
corresponding to the state which allows high contrast coherent
manipulation with $\eta^2 \bar{n} \ll 1$. Of course, the cooling
times are a function of the laser powers at 729~nm and 854~nm,
table~1 gives the values under typical operating conditions. As
can be seen from table~1, the characteristic times are all below
4~ms. This shows that with our chosen cooling time of 6~ms per
mode we have high ground state occupation probability in all modes
and this is largely sufficient to prevent spectator mode
decoherence of the coherent dynamics.

%heating rates
To yield the heating rate $\dot{\bar{n}}$ of one mode, we first
prepare the crystal in its motional ground state and then switch
off all laser beams.  We let the system evolve freely under the
influence of the  environment for a delay time $t_D$. Using the
same method as before we calculate the average phonon number in
the mode of interest and plot this value with the delay time, thus
we can measure the heating rate. Our measured heating rates for
various modes of the collective motion are summarized in table~2.
Typical heating rates in the Innsbruck Ca$^+$ traps are
$\approx$~10-20~phonons~s$^{-1}$. We have not observed an increase
of heating rates over a period of one year, unlike ref.
\cite{bigheat}, even though the linear trap was found to be
heavily coated with Calcium when the vacuum system was opened. In
addition, we did not observe a large difference between the
heating rates for c.o.m. and other modes.

\begin{table}[b]
\label{t_heat}
\begin{tabular}{l|cccccc}
Ion and Trap & Size $d$ & Mode & Frequency & Heating Rate  &
normalized Heating
 \\ & $\mu$m & & $(MHz)$ & (phonons/s) & (for 1~MHz) \\
 \hline
$^{40}Ca^+$, linear & 1180 &  breathing & 1.2 & 10 & 12 \\  & &
rocking & 1.7 & 8 & 14 \\   & & radial & 1.9 &   25(10) & 47(20)
\\

\\ $^{40}Ca^+$, spherical & 700 & c.o.m, axial & 4 & 5.2 & 21 \\ &
& c.o.m, radial & 1.9 & 14 & 27 \\

\\
$^{9}Be^+$, sph. \# 2 & 175 & c.o.m. (x) & 8.6 & 19 000
$^{+40000}_{13000}$ & 160 000$^{+340000}_{111000}$
\\ & & breathing-x & 15 & $\le$ 180 & $\le$ 2 700 \\
 &  & rocking-xy & 15 & $\le$ 1000 & $\le$ 15 000 \\

\\ $^{9}Be^+$, sph. \# 3b & 395 &  c.o.m. & 1.4
- 3.4 & varies & 5 000
\\
$^{9}Be^+$, lin. \# 4 & 280 & c.o.m. & 3 - 17 & " & 23 000 \\
$^{9}Be^+$, lin. \# 5 & 280 & c.o.m. & 3 - 10 & " & 35 000 \\
$^{9}Be^+$, lin. \# 6 & 365 &  c.o.m. & 3.5 - 10 & " & 11 000
\\

\\
 $^{198}Hg^+$, spherical & 450 & c.o.m & 3 & 6 & 18
\\
\end{tabular}
\caption{Heating rates for various traps and various motional
modes. For details on the spherical trap see
\cite{christian,stateeng}, while details on the Boulder traps for
Be$^+$ and Hg$^+$ ions are found in \cite{bigheat,cool2ions} and
\cite{single}.}
\end{table}

In order to compare the values of heating rates for the Ca$^+$
traps to other measured heating rates in spherical and linear
Be$^+$ traps \cite{bigheat}, we scale the motional heating rates
with the trap frequencies $\omega^{-1}$, and calculate the
normalized heating rate $\dot{N}$ at 1~MHz which is equivalent to
an absorbed power. As seen from table 2, both Ca$^+$ ion traps
show very low normalized heating rates. As discussed in ref.
\cite{bigheat}, the heating rate may well depend on the distance
$d$  between the electrode surface and the ion-crystal with the
fourth power $d^4$. For a discussion of the proposed
$d^4$-dependence, we scale the normalized heating rate with the
given distance and yield the proportionality coefficient for the
law $\dot{N} = c \cdot d^4[mm]$, which takes a value of
$c$~=~2.4(1.5) for the spherical Ca$^+$ Paul trap, 47(38) for the
linear Ca$^+$ trap and 164(38) for Be$^+$ if only c.o.m. modes are
averaged (120(83) for the overall average). Regarding the large
standard deviation of all values, it seems to be still quite
difficult to compare quantitatively the heating rates for
different traps and ions in order to identify a common mechanism.

\begin{figure}[b]
\begin{center}
\epsfig{file=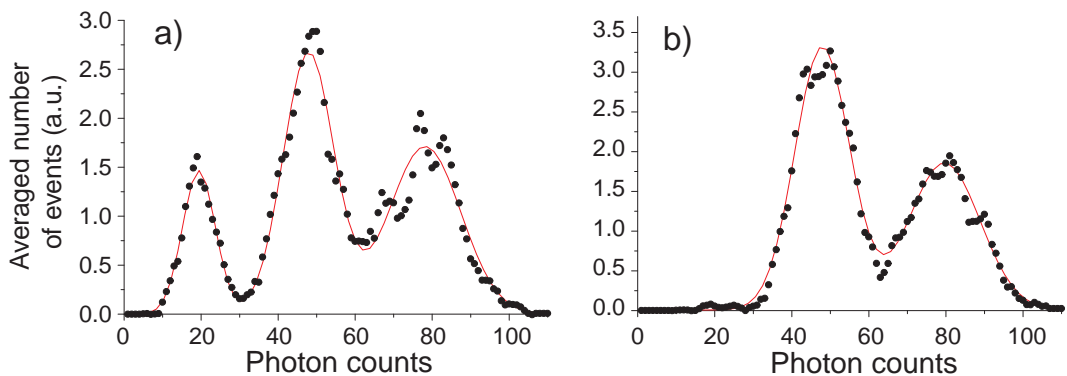, height=5cm} \caption{Both histograms are
the averaged sum of many 100 shot experiments. Histogram (a)
exhibits three peaks: one for no ionic excitation (far right
peak), one ion excited (middle) and both ions excited (far left).
Histogram (b) is collected with the laser addressing only one ion.
It is clear that the peak corresponding to two ion excitation has
disappeared.} \label{f_histo}
\end{center}
\end{figure}

\section{Sympathetic cooling\label{cooling}}
In this section we focus on the sympathetic nature of the ground
state cooling technique decribed above. For state detection
(section \ref{spec}, step (iv)) we use the electron shelving
technique, illuminating the complete ion string with light at
397~nm and 866~nm, and detecting the overall fluorescence. Ions in
the $S_{1/2}$ level scatter light on the dipole transition to the
$P_{1/2}$ state, while no fluorescence is observed if the ions
have been excited into the $D_{5/2}$ state. If we spatially adjust
the laser at 729~nm into the centre of the ion-crystal, we observe
a triple-peaked histogram as shown in fig.~\ref{f_histo}a. The
maxima near 19, 48 and 79 photons correspond to - {\em no ion} in
the $S_{1/2}$ state, background light only, - {\em one ion} of the
two in $S_{1/2}$ and - {\em both ions} in $S_{1/2}$ \cite{coat}. A
histogram is built up from many 100 shot experiments, each shot
having a detection time of 11.8~ms.

\begin{figure}
\begin{center}
\epsfig{file=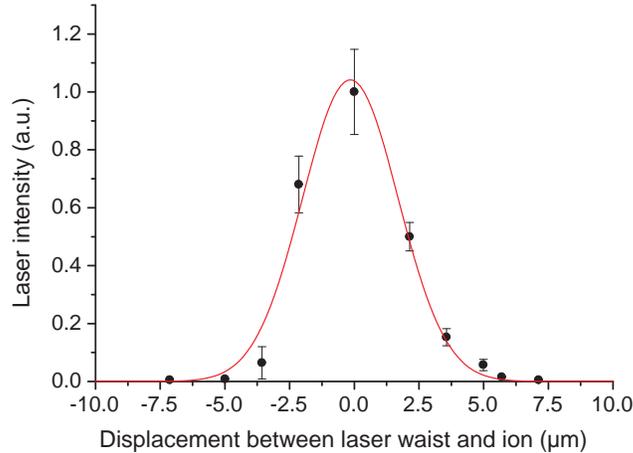, height=6cm} \caption{The width of the
laser beam, as measured by exciting a single ion in the trap and
observing Rabi oscillations. The waist, w$_{cool}$, is 3.7$\mu$m.}
\label{f_focus}
\end{center}
\end{figure}

If the 729~nm laser is illuminating only one of the two ions,
adjusted in a similar way as for the cooling of the
two-ion-crystal, we observe a double-peaked histogram. Now, only
the second maximum near 48 and the third near 79 photons are
visible (fig.~\ref{f_histo}b). With the laser focused on one of
the ions only, it is impossible to excite both ions into the
nonfluorescing $D_{5/2}$ state, therefore the first peak in the
histogram vanishes.

In a second measurement, with a single trapped ion in the linear
trap, we deduced the waist size of the addressing beam at 729~nm.
For this, we varied the position of the laser beam (in the axial
direction) with respect to the ion position and excited Rabi
oscillations on the $S_{1/2}$ - $D_{5/2}$ carrier transition. The
measured Rabi frequency maps the electric field amplitude directly
and we plot in fig.  \ref{f_focus} the data together with a
Gaussian fit which yields a 3.7(0.2)~$\mu$m waist size. For the
axial trap frequency of 700~kHz we calculate \cite{spectrosc} a
two-ion separation of 7.6~$\mu$m. The measured laser intensity
7~$\mu$m away from the centre (location of the non-illuminated
ion) is 5~$\cdot$~10$^{-3}$ in magnitude lower than that in the
centre of the addressing beam focus.

\begin{figure}
\begin{center}
\epsfig{file=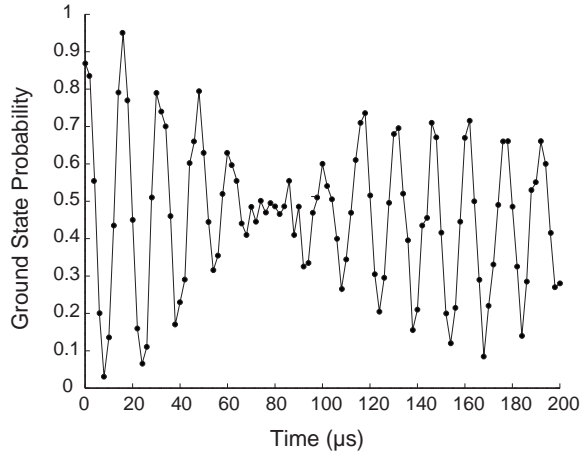, height=6cm} \caption{Rabi oscillations of
two ions. Experimental data are shown as points and have been
joined by lines to guide the eye. Note that the y axis is the
probablitiy of both ions being found in the non-excited state. The
collapse and revivals of the combined atomic population indicate
that we have created the states $\mid\uparrow\downarrow>$ and
$\mid\uparrow\uparrow>$ from the starting state
$\mid\downarrow\downarrow>$.} \label{f_rabi}
\end{center}
\end{figure}

\section{Coherent dynamics on the qubit transition}
With a two-ion-crystal cooled to the ground state as described in
section \ref{sideband}, we have performed coherent manipulation on
the qubit transition. Previously, in our group, we have cooled
single ions~\cite{stateeng,kuehtai} and demonstrated coherent
excitation on the qubit transition. Here we present results on the
electronic state of a two-ion-crystal. We excite the carrier of
the qubit transition after positioning the exciting laser
in-between the two ions rather than illuminating only one of them.
Rather than using a probe pulse of constant duration and power and
varying the frequency, we keep the frequency and power constant
and vary the duration. This way we build up a picture in time of
the coherent dynamics of the electronic states of each ion. In the
detection time, we need to set two thresholds in order to
distinguish between one and two ions having been excited. As can
be seen from the histogram (fig. \ref{f_histo}) these thresholds
can be placed accurately to distinguish the three possible states
of the combined atomic population when probed. The laser-ion
interaction, with the laser frequency tuned to the carrier
transition, will be described by different Rabi frequencies,
$\Omega_{Rabi}$, of each ion. This means that the population of
each ion will cycle with those different rates resulting in a
coherent transfer from the $S_{1/2}$ ground state and the
$D_{5/2}$ metastable state. These two frequencies $\Omega_{Rabi}$
will ''beat'' with each other when considering the combined atomic
population in the lower excited state. In this way, states such as
$\mid\uparrow\downarrow>$ are generated, where a
$\downarrow$($\uparrow$) indicates that an ion is in the
lower(upper) electronic state. If the spectator modes contained
many phonons it would not be possible to see this effect. With our
four sideband cooling and small Lamb-Dicke parameters we are able
to observe the single bit rotations (implemented on the two ions)
with high contrast. In fig.~\ref{f_rabi}, at $t=0$ both ions are
in the ground electronic state $\mid\downarrow\downarrow>$. At $t
\sim 75~\mu s$ the signal has 'collapsed' to 0.5. This means that
one ion has done an integer number of cycles up and down whereas
the other ion has undergone an extra half cycle up, thus we have
prepared the state $\mid\uparrow\downarrow>$. Now the reappearence
of the contrast occurs and at $t \sim 155~\mu s$ both ions have
been prepared in the upper electronic state,
$\mid\uparrow\uparrow>$. If there were no decoherence or heating
mechanisms, then the signal would not deteriorate, however laser
phase and magnetic field fluctuations cause the contrast in the
Rabi oscillations to be lost.

%sum up
\vspace{0.2cm} We have demonstrated sympathetic ground state
motional cooling of an ion chain with clear scaling possibilites
to longer chains. We have cooled the chain on all four motional
sidebands and acheived over 95$\%$ ground state populations in all
the motional modes. From these ground state ´populations, we have
measured the heating rates for various modes and found a typical
rate of the order of 10~phonons~s$^{-1}$. The measured heating
rates in our trap are very favourable for the preparation of the
motional ground state of an ion chain and its subsequent use as a
quantum information processor. This is in stark contrast to other
measurements where the normalized heating rate has been found a
factor of 1000 higher. By coherently exciting the qubit transition
we could demonstrate the preparation of product states of the two
ion chain. Our method of sympathetic motional ground state cooling
relies on the ability to individually mainpulate single ions which
is also necessary for many suggested QC and quantum error
correction schemes. The application of this technique will allow
us to generate entanglement and implement quantum gates between
ions in a chain.

This work is supported by the Austrian "Fonds zur F\"orderung der
wissenschaftlichen Forschung" within the project SFB15 and in
parts by the European Commission within the TMR networks "Quantum
Information" (ERB-FMRX-CT96-0087) and "Quantum Structures"
(ERB-FMRX-CT96-0077) and the "Institut f\"ur Quanteninformation
GmbH".

\hspace{0.5cm}


\begin{thebibliography}{99}
\bibitem{coat} After opening the vacuum chamber, we noticed the
photomultiplier observation viewport to be heavily coated with Ca
which reduced the overall count rate by at least a factor of 3.
\end{thebibliography}
\end{document}